\documentclass[runningheads]{llncs}

\usepackage[utf8]{inputenc}
\usepackage[T1]{fontenc}
\usepackage{time}
\usepackage[hyphens,allowmove]{url}

\usepackage{graphicx}

\usepackage[hidelinks]{hyperref}
\hypersetup{
  colorlinks   = true, 
  urlcolor     = blue, 
  linkcolor    = blue, 
  citecolor   = black 
}

\usepackage{color}
\usepackage{xcolor}

\newcommand{\red}[1]{\textbf{\textcolor{red}{#1}}}
\newcommand{\blue}[1]{\textbf{\textcolor{blue}{#1}}}

\newcommand{\purple}[1]{\textbf{{\color[HTML]{6600CC} #1}}}

\usepackage{fontawesome5}
\definecolor{idcolor}{HTML}{A6CE39}
\newcommand{\orcidlink}[1]{\href{https://orcid.org/#1}{\color{idcolor}\faOrcid}}

\usepackage{cite}
\usepackage{bookmark}

\def\figureautorefname~#1\null{Fig.~#1\null}

\usepackage{tabularray}
\UseTblrLibrary{booktabs}

\usepackage{ascmac}
\usepackage{amsmath,amssymb}

\usepackage{multirow}
\usepackage{colortbl}

\usepackage{cleveref}
\crefname{table}{Table}{Table}
\crefname{figure}{Fig.}{Fig.}
\crefname{section}{Section}{Section}

\newcommand{\Title}{Analysis of the deletions of DOIs}
\newcommand{\SubTitle}{What factors undermine their persistence and to what extent?}

\newcommand{\RQA}{How many deleted DOIs exist?}

\newcommand{\RQB}{Which document types are the most common in deleted DOIs?}

\newcommand{\RQC}{Which prefixes are the most common in deleted DOIs?}

\newcommand{\RQD}{What are the most common changes in the suffixes of deleted DOIs?}

\newcommand{\TopicA}{Basic Statistics of Deleted DOIs}
\newcommand{\TopicB}{Content Analysis Based on Crossref Metadata and DOI Links}
\newcommand{\TopicC}{Prefix Analysis}
\newcommand{\TopicD}{Suffix Analysis}

\hypersetup{
pdftitle={\Title \ - \SubTitle},
pdfauthor={Jiro Kikkawa, Masao Takaku, and Fuyuki Yoshikane}
}

\begin{document}

\title{\Title}
\subtitle{\SubTitle}

\author{Jiro Kikkawa\orcidlink{0000-0001-7258-7099} \and 
Masao Takaku\orcidlink{0000-0002-2458-6988} \and 
Fuyuki Yoshikane}

\authorrunning{J. Kikkawa et al.}

\institute{University of Tsukuba, Ibaraki, Japan \\
\email{\{jiro,masao,fuyuki\}@slis.tsukuba.ac.jp}}

\maketitle

\begin{abstract}
Digital Object Identifiers (DOIs) are regarded as persistent; however, they are sometimes deleted. Deleted DOIs are an important issue not only for persistent access to scholarly content but also for bibliometrics, because they may cause problems in correctly identifying scholarly articles. However, little is known about how much of deleted DOIs and what causes them. We identified deleted DOIs by comparing the datasets of all Crossref DOIs on two different dates, investigated the number of deleted DOIs in the scholarly content along with the corresponding document types, and analyzed the factors that cause deleted DOIs. Using the proposed method, 708,282 deleted DOIs were identified. The majority corresponded to individual scholarly articles such as journal articles, proceedings articles, and book chapters. There were cases of many DOIs assigned to the same content, e.g., retracted journal articles and abstracts of international conferences. We show the publishers and academic societies which are the most common in deleted DOIs. In addition, the top cases of single scholarly content with a large number of deleted DOIs were revealed. The findings of this study are useful for citation analysis and altmetrics, as well as for avoiding deleted DOIs.

\keywords{Digital Object Identifier (DOI) \and Scholarly Communication \and Bibliometrics \and Altmetrics.}
\end{abstract}

\section{Introduction}

It is important to ensure persistent access to scholarly articles because many scholarly articles have been distributed on the Web. Digital Object Identifier (DOI) is the best-known persistent identifier of scholarly content and is an international standard. As of May 2022, 130 million DOIs have been registered by Crossref~\cite{Crossref_stats}.

In the DOI system, a single and unique DOI is registered for each content item. A DOI name consists of a prefix, slash (/), and suffix. A prefix is assigned to a particular DOI registrant, such as publishing companies or academic societies. DOI registrants assign suffixes to their content and register DOIs through DOI registration agencies (hereinafter referred to as ``RAs''). DOIs also provide hyperlinks (hereinafter referred to as ``DOI links'') by adding DOI name after ``\url{https://doi.org/}.'' Content holders, such as publishers and academic societies, maintain the relationship between the DOI and the original content's URI. Through these functions, scholarly content can be consistently accessed using DOI links. Once registered, the DOIs are regarded as persistent and are not deleted.

Because of these characteristics of DOIs, citation indexes and altmetrics depend on them to identify unique scholarly articles. With the rapid increase in the number of scholarly articles, DOIs are playing an increasingly significant role in scholarly communication. However, in practice, DOIs and their artifacts (e.g., identifiers, metadata, and redirected URIs) may be deleted by content holders, such as publishers and academic societies, if something wrong happens, e.g., multiple DOIs were assigned to the same content item. These deletions cause problems not only for persistent access to scholarly content but also for bibliometric analysis, including citation analysis and altmetrics. 
In this study, we focus on the deletion of identifiers and metadata, and define DOIs for which identifiers and associated metadata cannot be retrieved as ``Deleted DOIs.''

Deleted DOIs are an important issue in bibliometrics; however, little is known about their quantity and causes. Thus, we developed a methodology to identify deleted DOIs and analyze not only their quantity but also their causes, with a focus on the content holders and their patterns of DOI assignment.

The following research questions were addressed in this study:

\begin{description}
	\item[RQ1] \RQA
 \begin{itemize}
	\item We present the total number of deleted DOIs and classify deleted DOIs according to the redirect URIs and Crossref metadata.
 \end{itemize}
	\item[RQ2] \RQB
	\begin{itemize}
	\item We investigate which document types have large numbers of deleted DOIs; i.e., we clarify whether these content items are single scholarly articles (such as a journal article, conference proceedings paper, or book chapter) or non-single scholarly articles (such as an entire book or a specific volume or issue of a journal). If a large number of deleted DOIs were registered for a single scholarly article, we analyze them in detail according to their metadata.
 \end{itemize}
	\item[RQ3] \RQC
 \begin{itemize}
	\item We find publishers and academic societies with large numbers of deleted DOIs.
 \end{itemize}
	\item[RQ4] \RQD
 \begin{itemize}
	\item We investigate the pattern of changes in the suffixes of DOIs for the same scholarly content to determine the relationship between DOI assignment patterns and deleted DOIs.
 \end{itemize}
\end{description}

The contributions of this study are twofold. (1) We present the overall picture of deleted DOIs by clarifying the number of deleted DOIs and the characteristics of their content. The potential impact of deleted DOIs for citation analysis and altmetrics is demonstrated. (2) By identifying the factors that cause deleted DOIs, we provide guidance for avoiding deleted DOIs and making the DOI system more stable.

\section{Related Work}

\subsubsection{Crossref DOI statistics.}
Hendricks et al.~\cite{Hendricks_Tkaczyk_Lin_Feeney_2020} reported the statistics of Crossref DOIs in June 2019. More than 106 million Crossref DOIs had been registered, and the number of DOIs had increased by 11\% on average over the past 10 years. As for the types of contents, 73\% are journals, 13\% are books, and 5.5\% are conference papers and proceedings.

\vspace{-0.4cm}
\subsubsection{Investigation of duplicated Crossref DOIs.}
Tkaczyk~\cite{Crossref_Double_trouble_with_DOIs} investigated Crossref DOIs not marked as an alias to other DOIs to consider their quantity and impact on citation-based metrics. Among DOIs randomly sampled from 590 publishers and academic societies with $\geqq5,000$ DOIs, 0.8\% were duplicated, i.e., different DOI names but their metadata were the same or highly similar. The majority of them were caused by the re-registration of DOIs by the same publishers and academic societies. As for duplicated DOIs among different publishers and academic societies, one of the most frequent cases was content with DOIs initially registered by JSTOR and re-registered by new content holders.

\vspace{-0.4cm}
\subsubsection{Incorrect DOIs indexed by scholarly bibliographic databases.}
\label{analyses_of_the_dois_indexed_by_the_databases}

Several studies have revealed errors in DOIs indexed by scholarly bibliographic databases. Franceschini et al.~\cite{Franceschini_Maisano_Mastrogiacomo_2015} analyzed DOIs in the records of Scopus and found that multiple DOIs were incorrectly assigned to the same record as rare cases. Zhu et al.~\cite{Zhu_Hu_Liu_2019} analyzed DOIs in the Web of Science records. They reported not only ``wrong DOI names'' but also ``one paper with two different DOI names.'' The former are similar errors, as reported by Franceschini et al.~\cite{Franceschini_Maisano_Mastrogiacomo_2015}.
The latter are classified into the following two cases: (1) there were both correct and incorrect DOIs in the records; (2) multiple correct DOIs were assigned to the same scholarly article.

\vspace{-0.4cm}
\subsubsection{Analysis of persistence of Crossref DOIs.}
\label{analysis_by_klein}

Klein and Balakireva~\cite{Klein_Balakireva_2020,Klein_Balakireva_2021} examined the persistence of Crossref DOIs by analyzing their HTTP status codes. They randomly extracted 10,000 Crossref DOIs and examined the final status codes for each DOI link by using multiple HTTP request methods. More than half of the DOI links did not redirect to the content when an external network from academic institutions was used. However, the errors of all the DOI links were reduced to one-third when an internal network from academic institutions was used. These results indicate that the responses for the same DOI can differ according to conditions such as the HTTP request methods and network locations, which implies a lack of persistence of DOIs.

\vspace{-0.4cm}
\subsubsection{Analysis of the usage of DOI links in scholarly articles.}
Regarding the usage of DOI links in the references of scholarly references, Van de Sompel et al.~\cite{Van_de_Sompel2016} examined references from 1.8 million papers published between 1997 and 2012. Consequently, they identified a problem that numerous scholarly articles were referenced using their location URIs instead of their DOI links. \\

As described previously, researchers have reported duplicated Crossref DOIs~\cite{Crossref_Double_trouble_with_DOIs,Zhu_Hu_Liu_2019}, and some Crossref DOIs cause errors and are unable to lead to the contents~\cite{Klein_Balakireva_2020,Klein_Balakireva_2021}. It has also been reported that DOIs included in the records of scholarly bibliographic databases contain errors~\cite{Franceschini_Maisano_Mastrogiacomo_2015,Zhu_Hu_Liu_2019}. However, these studies were based on relatively small samples because a methodology to identify duplicated or deleted DOIs at a large scale has not been proposed. Thus, we propose a methodology for identifying deleted DOIs and conducting a large-scale analysis to determine how many deleted DOIs exist and what factors cause them.

\section{Materials and Methods}

\subsection{Identifying Deleted DOIs}
\label{target_filtering}

In this section, we describe the proposed method for identifying deleted DOIs, i.e., DOIs whose identifiers and metadata cannot be retrieved. Specifically, we extracted the DOIs that existed at a point in time and did not exist later by comparing the dump files of Crossref DOIs on two different dates. We then treated these DOIs as a candidate set of deleted DOIs.

First, we obtained the two datasets of all Crossref DOIs on different dates: March 2017~\cite{Himmelstein_Wheeler_Greene_2017} and January 2021~\cite{Kemp,CrossrefPublicDataFile2021}. We unescaped and downcased to remove duplicates. The total numbers of unique DOIs were 87,542,203 and 120,684,617, respectively.

Next, we obtained the differences and product sets between the two datasets, as shown in \cref{tbl:basic_stats}. There were 711,198 Crossref DOIs that existed as of March 2017 but did not exist as of January 2021. Then, for the 711,198 DOIs, we (1) identified the RA, (2) examined the redirected URI of DOI links, and (3) obtained the Crossref metadata. For (1), we used a web API called ``Which RA?''~\cite{doiRA_doc}, which returns the RA name if the DOI exists and returns an error if it does not exist. For (2), we examined the HTTP header of each DOI link. For (3), we used the Crossref REST API~\cite{crossref_rest_api,crossref_rest_api_format}.

\begin{table*}[htb]
	\begin{center}
		\caption{Basic statistics of the datasets}
		\label{tbl:basic_stats}
		\begin{tblr}{l|rr|rr}
			\toprule
			& \SetCell[c=2]{c} \textbf{Dataset as of March 2017} &
			& \SetCell[c=2]{c} \textbf{Dataset as of January 2021} & \\

			& \textbf{\# of DOIs} & \textbf{\%} & \textbf{\# of DOIs} & \textbf{\%} \\
			\midrule[\heavyrulewidth]
			\textbf{Difference set} & \textbf{711,198} & \textbf{0.81} & 33,853,612 & 28.05 \\
			\textbf{Product set} & 86,831,005 & 99.19 & 86,831,005 & 71.95 \\
			\textbf{Overall} & 87,542,203 & 100.00 & 120,684,617 & 100.00 \\
			\bottomrule
		\end{tblr}
	\end{center}
\end{table*}

\begin{figure*}[htb]
	\centering
	\includegraphics[width=\linewidth]{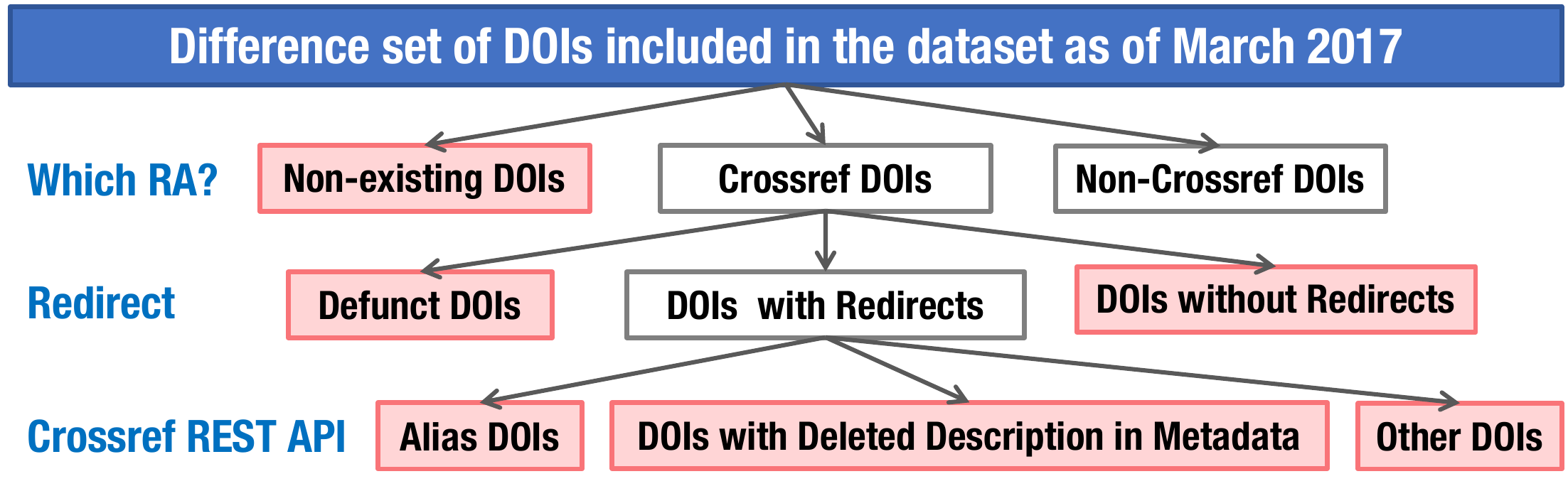}
	\caption{Classification of the DOIs that existed as of March 2017 but no longer existed as of January 2021. Items with a red box correspond to deleted DOIs.}
	\label{fig:doi_research_target_overview}
\end{figure*}

According to the results for (1) to (3), we classified the DOIs into the groups shown in \cref{fig:doi_research_target_overview}. First, we classified them into ``Non-existing DOIs,'' ``Crossref DOIs,'' and ``Non-Crossref DOIs'' according to the results of ``Which RA?''; these groups were derived from the results of ``DOI does not exist,'' ``Crossref,'' and RA names other than Crossref, respectively. Among these groups, Non-existing DOIs corresponded to deleted DOIs. In contrast, Non-Crossref DOIs were not deleted, because they were not included in the dataset as of March 2021, owing to the transfer of the RAs.

Next, we classified the Crossref DOIs into the groups ``Defunct DOIs,'' ``DOIs without Redirects,'' and ``DOIs with Redirects'' according to the results of each HTTP header for the DOI links. ``Defunct DOIs'' were the DOIs that redirected to the specific URI (\url{https://www.crossref.org/_deleted-doi/}) for the deleted content~\cite{Defunct_DOI}. ``DOIs without Redirects'' were the DOIs without redirects when we accessed DOI links; thus, we could not reach the content via these DOIs. We treated defunct DOIs and DOIs without redirects as deleted DOIs.

Finally, the DOIs were classified into three groups according to the results of the Crossref REST API. They were all deleted DOIs, because they were unable to retrieve the Crossref metadata.

\begin{itemize}
	\item \textbf{Alias DOI}: When multiple Crossref DOIs were assigned to the same content item, one was set as the Primary DOI, and the others were set as the Alias DOIs~\cite{Conflict_report}. The Crossref REST API returned the error ``Resource not found'' for the Alias DOIs.
	\item \textbf{DOI with Deleted Description in Metadata}: Crossref metadata for a particular DOI or the corresponding Alias DOI contains the word ``delete'' in the value of \ ``title'' or ``container title.'' Moreover, the first author manually confirmed that the content was deleted by accessing the landing page.
	\item \textbf{Other DOI}: These DOIs had redirects that were not applicable to any of the conditions above.
\end{itemize}

When the DOIs were applicable to both the ``Alias DOI'' and ``DOI with Deleted Description in Metadata'' groups, we assigned them to the latter group.

\subsection{Analysis Methods}
\label{methods}

The following four analyses were performed on the deleted DOIs identified by the method described in \cref{target_filtering}. Each analysis corresponds to an RQ.

\vspace{-0.3cm}
\subsubsection{(1) \TopicA.}
We investigated the number of deleted DOIs and their distributions using the groups presented in \cref*{target_filtering}.

\vspace{-0.3cm}
\subsubsection{(2) \TopicB.}
First, we determined which content types were common among the deleted DOIs according to the value of ``type'' in the Crossref metadata. Next, we investigated the number of alias DOIs corresponding to a single Primary DOI to clarify the characteristics of deletion when multiple DOIs were registered to the same content. For example, when Alias DOIs ``10.14359/15303,'' ``10.14359/15304,'' and ``10.14359/15305'' corresponded to the same Primary DOI ``10.14359/15306,'' the number of Alias DOIs was 3. We obtained the basic statistics of the Alias DOIs corresponding to the same Primary DOIs. We examined articles for cases in which many Alias DOIs were associated with a single Primary DOI.

\vspace{-0.3cm}
\subsubsection{(3) \TopicC.}
\label{Method_TopicC}

We compared the prefixes of deleted DOIs and identified which publishers or academic societies had large numbers of deleted DOIs. In addition, we calculated the ratios $P_1$ and $P_2$ for each prefix, as follows.

\begin{eqnarray}
	P_1 (X) & = & \frac{total \ number \ of \ the \ Deleted \ DOIs \ whose \ Prefix \ was \ X}{total \ number \ of \ all \ Deleted \ DOIs} * 100 \nonumber
\end{eqnarray}
\begin{eqnarray}
	P_2 (X) & = & \frac{total \ number \ of \ the \ Deleted \ DOIs \ whose \ Prefix \ was \ X}{total \ number \ of \ all \ DOIs \ whose \ Prefix \ was \ X \ as \ of \ March \ 2017} * 100 \nonumber
\end{eqnarray}

When both $P_1$ and $P_2$ are high in a prefix, it indicates not only that the percentage of deleted DOIs for this prefix in the total deleted DOIs is high but also that a large percentage of the DOIs registered by the registrant with this prefix are deleted. On the other hand, when $P_1$ is high and $P_2$ is low in a prefix, the percentage of deleted DOIs in the total deleted DOIs is high but a small percentage of the DOIs registered by the registrant with that prefix are deleted.

\vspace{-0.3cm}
\subsubsection{(4) \TopicD.}

We analyzed the suffixes of the deleted DOIs when multiple DOIs were registered for the same content. Here, we set the target to the pairs of Alias DOIs and Primary DOIs. First, we classified the patterns of the pairs of Alias DOI and Primary DOI suffixes into the following categories: (A) only the suffix changed (B) both the prefix and the suffix changed, and (C) only the prefix changed. Next, for (A) and (B), we analyzed the similarity of strings between the suffixes of the Alias DOI and Primary DOI. The Levenshtein distance~\cite{Levenshtein1966} normalized by the string length of the suffixes was used as the similarity score. The following formula was used:

\vspace{-0.4cm}
\begin{eqnarray}
	Sim(str1, str2) & = & 1- \frac{Levenshtein(str1, str2)}{Max(|str1|,|str2|) \ \ ,} \nonumber
\end{eqnarray}

where $Sim$ takes values between 0 and 1, with higher values indicating a higher degree of similarity between the two strings. We observed the distributions for the similarity score in groups (A) and (B). When the similarity scores of the majority in a group are high, it can be interpreted that there are many cases where slightly changed suffixes of Alias DOIs are set as those of Primary DOIs. In contrast, when the similarity score of the majority in a group is low, it can be interpreted that there are many cases where strings that differ significantly from the Alias DOI suffixes are set as the suffixes of Primary DOIs.

Finally, we analyzed the character-level differences between each pair of Alias and Primary DOIs to determine what types of changes were common. We used the package ``diff-lcs''~\cite{diff_lcs} from RubyGems to obtain the diffs of the Primary and Alias DOIs using the Diff::LCS.diff method. We retrieved the changes between each Primary DOI and Alias DOI as actions of adding or deleting the position and the corresponding characters. We analyzed which characters were added, deleted, or replaced (when both ``delete'' and ``add'' were present in the same position) for each pair. For example, in the case of the Alias DOI ``10.14359/1530\textbf{3}'' and Primary DOI ``10.14359/1530\textbf{6},'' the change was to replace ``3'' with ``6'' once.

\section{Results and Discussion}

\subsection{\TopicA}

\begin{description}
	\item[RQ1] \RQA
\end{description}

\cref{tbl:doi_research_target_overview} presents the number of deleted DOIs in each group. We identified 708,282 deleted DOIs by applying the method described in Section 3.1. Most deleted DOIs were Alias DOIs, accounting for 94.29\% of the total. Therefore, the majority of the deleted DOIs were caused by the deletion of multiple DOI assignments for the same content.
The percentage of deleted DOIs relative to all DOIs in 2021 was 0.59\% ($=708,282/120,684,617*100$).

\begin{table}[htb]
	\begin{center}
		\caption{Number of deleted DOIs in each group (n=708,282)}
		\label{tbl:doi_research_target_overview}
		\begin{booktabs}[\textwidth]{@{\extracolsep{\fill}} rlrr @{\extracolsep{\fill}}}
			\toprule
			\# & \textbf{Group} & \textbf{Count} & \textbf{\%} \\
			\midrule
			1 & Non-existing DOIs & 240 & 0.03 \\
			2 & Defunct DOIs & 388 & 0.05 \\			
			3 & DOIs without redirects & 693 & 0.10 \\
			4 & Alias DOIs & 667,869 & 94.29 \\
			5 & DOIs with deleted description on metadata & 1,144 & 0.16 \\
			6 & Other DOIs & 37,948 & 5.36 \\
			\bottomrule
		\end{booktabs}
	\end{center}
\end{table}

\vspace{-0.8cm}
\subsection{\TopicB}

\begin{description}
	\item[RQ2] \RQB
\end{description}

\cref{tbl:count_by_type} presents the document types of the deleted DOIs based on Crossref metadata. The top three types were journal articles (78.35\%), proceedings articles (8.48\%), and book chapters (6.77\%). Thus, most of the deleted DOIs were individual scholarly articles rather than books or journals.

\begin{table}[htb]
	\begin{center}
		\caption{Document types of deleted DOIs (n=708,282)}
		\label{tbl:count_by_type}
		\begin{booktabs}[\textwidth]{@{\extracolsep{\fill}} rlrr|rlrr @{\extracolsep{\fill}}}
			\toprule
			\# & \textbf{Type} & \textbf{Count} & \textbf{\%} & \# & \textbf{Type} & \textbf{Count} & \textbf{\%} \\
			\midrule
			1 & journal-article & 554,948 & 78.35 & 10 & standard & 70 & 0.01 \\
			2 & proceedings-article & 60,092 & 8.48 & 11 & book-series & 48 & 0.01 \\
			3 & book-chapter & 47,974 & 6.77 & 12 & report & 40 & 0.01 \\
			4 & dataset & 23,784 & 3.36 & 13 & journal-issue & 37 & 0.01 \\
			5 & component & 8,419 & 1.19 & 14 & proceedings & 23 & 0.00 \\
			6 & reference-entry & 6,611 & 0.93 & 15 & reference-book & 21 & 0.00 \\
			7 & book & 3,094 & 0.44 & 16 & journal & 15 & 0.00 \\
			8 & monograph & 2,340 & 0.33 & 17 & dissertation & 10 & 0.00 \\
			9 & other & 754 & 0.11 & 18 & book-section & 2 & 0.00 \\
			\bottomrule
		\end{booktabs}
	\end{center}
\end{table}

\begin{table}[htb]
	\begin{center}
		\caption{Ten Primary DOIs with the largest numbers of associated Alias DOIs}
		\vspace{-0.4cm}
		\label{tbl:pairs_of_alias_primay_top_list}
		\begin{booktabs}[\textwidth]{@{\extracolsep{\fill}} rll @{\extracolsep{\fill}}}
			\toprule
			\# & \textbf{Primary DOI} & \textbf{Title} \\
			\textbf{Count} & \textbf{Volume (Issue), Page} & \textbf{Container title} \\
			\midrule
			1 & \href{https://doi.org/10.1016/s1876-6102(14)00454-8}{10.1016/s1876-6102(14)00454-8} & Volume Removed - Publisher's Disclaimer \\
			1,474 & 13, pp.~1--10380 & Energy Procedia \\ \midrule
			2 & \href{https://doi.org/10.1016/s1876-6102(14)00453-6}{10.1016/s1876-6102(14)00453-6} & Volume Removed - Publisher's Disclaimer \\
			748 & 11, pp.~1--5156 & Energy Procedia \\ \midrule

			3 & \href{https://doi.org/10.1016/j.fueleneab.2006.10.002}{10.1016/j.fueleneab.2006.10.002} & Abstracts \\
			486 & 47~(6), pp.~384--446 & Fuel and Energy Abstracts \\ \midrule

			4 & \href{https://doi.org/10.1016/s0735-1097(01)80004-8}{10.1016/s0735-1097(01)80004-8} & Hypertension, vascular disease, and prevention \\ 399 & 37~(2), pp.~A220--A304 & Journal of the American College of Cardiology \\ \midrule

			5 & \href{https://doi.org/10.1016/s0735-1097(01)80001-2}{10.1016/s0735-1097(01)80001-2} & ACCIS2001 angiography \& interventional ... \\ 394 & 37~(2), pp.~A1--A86 & Journal of the American College of Cardiology \\ \midrule

			6 & \href{https://doi.org/10.1016/s0735-1097(01)80003-6}{10.1016/s0735-1097(01)80003-6} & Cardiac function and heart failure \\ 356 & 37~(2), pp.~A142--A219 & Journal of the American College of Cardiology \\ \midrule

			7 & \href{https://doi.org/10.1016/s0735-1097(01)80005-x}{10.1016/s0735-1097(01)80005-x} & Myocardial ischemia and infarction \\
			333 & 37~(2), pp.~A305--A378 & Journal of the American College of Cardiology \\ \midrule

			8 & \href{https://doi.org/10.1016/s0735-1097(01)80006-1}{10.1016/s0735-1097(01)80006-1} & Noninvasive imaging \\
			321 & 37 (2), pp.~A379--A452 & Journal of the American College of Cardiology \\ \midrule

			9 & \href{https://doi.org/10.1016/s0735-1097(01)80002-4}{10.1016/s0735-1097(01)80002-4} & Cardiac arrhythmias \\
			254 & 37~(2), pp.~A87--A141 & Journal of the American College of Cardiology \\ \midrule

			10 & \href{https://doi.org/10.1037/h0099051}{10.1037/h0099051} & Forty-Third Annual Meeting of the American ... \\ 162 & 36~(2), pp.~195--373 & American Journal of Orthopsychiatry \\
			\bottomrule




		\end{booktabs}
	\end{center}
\end{table}

The minimum value, maximum value, median, and standard deviation of the number of alias DOIs corresponding to each primary DOI were 1, 1,474, 1, and 2.49, respectively. The majority of cases were those in which two DOIs were registered for the same content, and one of them was deleted because the median value was 1. On the other hand, focusing on the maximum value, there were cases where a large number of DOIs were assigned to a single content item.

\cref{tbl:pairs_of_alias_primay_top_list} presents the 10 Primary DOIs with the largest numbers of associated Alias DOIs. The content item with the largest number of associated Alias DOIs was Volume 13 of the journal \textit{Energy Procedia} (\#1 in \cref*{tbl:pairs_of_alias_primay_top_list}).
This entire volume was retracted. These articles are different scholarly articles, and it is considered that the withdrawn information should be applied to the respective DOIs, rather than applying them as alias DOIs to a single primary DOI. In case \#2, all articles published in Volume 11 of \textit{Energy Procedia} were withdrawn. Case \#3 involved abstracts of an international conference, where DOIs were registered to the abstracts of the presentations, but the publisher appeared to change the policy for registering DOIs to a set of abstracts. Cases \#4--\#9 were similar to \#3 and all involved the \textit{Journal of the American College of Cardiology}. Case \#10 involved an abstract of an international conference published in another journal. According to these results, the Primary DOIs with large numbers of Alias DOIs corresponded to content such as retracted journal articles and abstracts of international conferences. This content would not be the target of citation analysis and altmetrics and did not have a critical impact on them. The Primary DOIs with large numbers of Alias DOIs were caused by the activities of publishers and academic societies rather than by the deletion of multiple DOIs mistakenly registered to the same scholarly articles.

\subsection{\TopicC}

\begin{description}
	\item[RQ3] \RQC
\end{description}

\begin{table}[htb]
	\begin{center}
		\caption{Top 10 prefixes of deleted DOIs (n=708,282)}
		\vspace{-0.4cm}
		\centering
		\label{tbl:count_deleted_doi_prefix}
		\begin{booktabs}[\textwidth]{@{\extracolsep{\fill}} rllrrr @{\extracolsep{\fill}}}
			\toprule
			\# & \textbf{Prefix} & \textbf{Registrant} & \textbf{Count} & \textbf{$P_1$} (\%) & \textbf{$P_2$} (\%) \\
			\midrule[\heavyrulewidth]
			1 & 10.1002 & Wiley & 94,471 & 13.34 & 2.00  \\
			2 & 10.1016 & Elsevier BV & 85,643 & 12.09 & 0.58 \\
			3 & 10.2307 & JSTOR & 66,232 & 9.35 & 2.82 \\
			4 & 10.1037 & American Psychological Association & 51,846 & 7.32 & 6.74 \\
			5 & 10.2523 & Society of Petroleum Engineers & 35,499 & 5.01 & 87.62 \\
			6 & 10.1163 & Brill & 23,962 & 3.38 & 2.68 \\
			7 & 10.4018 & IGI Global & 23,440 & 3.31 & 17.51 \\
			8 & 10.1080 & Informa UK Limited & 21,134 & 2.98 & 0.75 \\
			9 & 10.1007 & Springer Science and Business Media LLC & 19,807 & 2.80 & 0.21 \\
			10 & 10.1111 & Wiley & 18,743 & 2.65 & 0.64 \\
			\bottomrule
		\end{booktabs}
	\end{center}
	\vspace{-0.1cm}
\end{table}

\cref{tbl:count_deleted_doi_prefix} presents the top 10 prefixes of the deleted DOIs. $P_1$ and $P_2$ in \cref*{tbl:count_deleted_doi_prefix} are the two ratios defined in \cref{Method_TopicC}. For example, the publisher Wiley registered 4,718,360 Crossref DOIs with the prefix ``10.1002'' as of March 2017, and 94,471 of them were Deleted DOIs. In this case, $P_1$ was $13.34\% (=94,471/708,282*100)$, and $P_2$ was $2.00\% (=94,471/4,718,360*100)$. As indicated by \cref*{tbl:count_deleted_doi_prefix}, the registrants with the largest numbers of deleted DOIs were major international commercial publishers, such as Wiley, Elsevier BV, and Springer Science and Business Media LLC (\#1, \#2, \#9, and \#10). These publishers registered numerous DOIs, but few of them were deleted, as the $P_2$ values for these publishers were low. In contrast, the ``Society of Petroleum Engineers'' (\#5) had a remarkably high $P_2$ value of 87.62\%; thus, most of the DOIs registered by this registrant with this prefix were deleted. IGI Global and the American Psychological Association (\#7 and \#4) had relatively high $P_2$ values of 17.51\% and 6.74\%, respectively. JSTOR (\#3)---a digital library platform---was a registrant with many deleted DOIs. 
Regarding JSTOR, the deletions of DOI seem to be caused by the re-registration of DOIs to the same article published by content holders other than JSTOR. For these cases, it is necessary to assign a single DOI to the initial version of the article and avoid re-registration of the DOIs to the other versions. This problem would be caused by the registration agency improperly managing their content.

\subsection{\TopicD}

\begin{description}
	\item[RQ4] \RQD
\end{description}

\begin{table}[htb]
	\begin{center}
		\caption{DOI name change patterns (n=667,869)}
		\centering
		\label{tbl:doi_name_change_patterns}
		\begin{booktabs}[\textwidth]{@{\extracolsep{\fill}} lrr @{\extracolsep{\fill}}}
			\toprule
			\textbf{Pattern} & \textbf{Count} & \textbf{\%} \\
			\midrule[\heavyrulewidth]
			Only the suffix changed & 465,789 & 69.74 \\
			Both the prefix and the suffix changed & 156,650 & 23.46 \\
			Only the prefix changed & 45,430 & 6.80 \\
			\bottomrule
		\end{booktabs}
	\end{center}
\end{table}

\cref{tbl:doi_name_change_patterns} presents the DOI name change patterns for each pair of Alias and Primary DOIs. The most common pattern was ``only the suffix changed'' (69.74\%), indicating that many DOIs registered under the same prefix were deleted when multiple DOIs were registered to the same content by the same registrant. In contrast, ``both the prefix and the suffix changed'' and ``only the prefix changed'' were cases of deletions in which the DOIs were re-registered with the content holder's prefix when the content holder changed. For the prefixes ``10.4018 (IGI Global)'' and ``10.1037 (American Psychological Association)'' (\#7 and \#4, respectively, in \cref*{tbl:count_deleted_doi_prefix}), almost all the patterns were ``only the suffix changes.''

Of the 156,650 cases in which both the prefix and the suffix changed, 56,491 were caused by the prefix ``10.2307 (JSTOR).'' The prefix ``10.2307'' was changed to ``10.1090 (American Mathematical Society)'' and ``10.1080 (Informa UK Limited)'' in 22,198 and 10,010 cases, respectively. These results match those of Tkaczyk~\cite{Crossref_Double_trouble_with_DOIs} and indicate that a large amount of scholarly content provided by JSTOR was transferred to other registrants, and their DOIs were re-registered with the prefixes of new registrants, which is why the prefix ``10.2307'' is listed as \#3 in \cref*{tbl:count_deleted_doi_prefix}.

Of the 45,430 cases of ``only the prefix changed,'' 35,176 involved the prefix ``10.2523'' of Alias DOIs and the prefix ``10.2118'' of Primary DOIs, and the corresponding registrant was the Society of Petroleum Engineers. This is why the prefix ``10.2523'' is listed as \#5 in \cref*{tbl:count_deleted_doi_prefix}, and these cases were due to DOIs being re-registered by the same publisher.

\begin{table}[htb]
	\begin{center}
		\caption{Distribution of similarity scores between Alias DOI and Primary DOI suffixes}
		\label{tbl:list_similarity_score}
		\begin{tblr}{l|rr|rr}
			\toprule
			& \SetCell[c=2]{c} \textbf{Only the suffix changed} &
			& \SetCell[c=2]{c} \textbf{Both the prefix and} & \\
			& \SetCell[c=2]{c} \textbf{} &
			& \SetCell[c=2]{c} \textbf{the suffix changed} & \\

			& \textbf{Count} & \textbf{\%} & \textbf{Count} & \textbf{\%} \\
			\midrule[\heavyrulewidth]
			$0 \leqq sim \leqq 0.1$ & 9,033 & 1.94 & \textbf{15,942} & \textbf{10.18} \\
			$0.1 < sim \leqq 0.2$ & 35,253 & 7.57 & \textbf{80,420} & \textbf{51.34} \\
			$0.2 < sim \leqq 0.3$ & 20,058 & 4.31 & \textbf{34,385} & \textbf{21.95} \\
			$0.3 < sim \leqq 0.4$ & 9,539 & 2.05 & 13,278 & 8.48 \\
			$0.4 < sim \leqq 0.5$ & 27,074 & 5.81 & 6,493 & 4.14 \\
			$0.5 < sim \leqq 0.6$ & 27,783 & 5.96 & 2,243 & 1.43 \\
			$0.6 < sim \leqq 0.7$ & 26,331 & 5.65 & 878 & 0.56 \\
			$0.7 < sim \leqq 0.8$ & \textbf{57,866} & \textbf{12.42} & 536 & 0.34 \\
			$0.8 < sim \leqq 0.9$ & \textbf{82,647} & \textbf{17.74} & 2,361 & 1.51 \\
			$0.9 < sim < 1.0$ & \textbf{170,205} & \textbf{36.54} & 114 & 0.07 \\
			\midrule[\heavyrulewidth]
			Overall & 465,789 & 100.00 & 156,650 & 100.00 \\
			\bottomrule
		\end{tblr}
	\end{center}
\end{table}

\cref{tbl:list_similarity_score} presents the distribution of similarity scores among suffixes for ``only the suffix changed'' and ``both the prefix and the suffix changed.'' The group ``only the suffix changed'' had a large percentage of cases with $Sim$ values of $>0.7$, indicating that most of the cases had a high degree of similarity among suffixes. In contrast, the group ``both the prefix and the suffix changed'' had a large percentage of cases with Sim values of $\leqq0.3$, indicating that the degree of similarity among the suffixes was low. This suggests that a slightly changed version of the suffix of the Alias DOI tended to be set as the suffix of the Primary DOI for the group ``only the suffix changed.'' In contrast, the group ``both the prefix and the suffix changed'' have a tendency that a significantly different suffix from the Alias DOI was set as the suffix of the Primary DOI.

\begin{table}[htb]
	\begin{center}
		\caption{Three most frequent patterns for ``only the suffix changed'' (n=465,789). Red, blue, and purple text corresponds to deletion, addition, and replacement, respectively.}
		\centering
		\label{tbl:diff_list_only_suffix_changes}
		\begin{booktabs}[\textwidth]{rlrr @{\extracolsep{\fill}}}
			\toprule
			\# & \textbf{Pattern of changes in the suffix} & \textbf{Count} & \textbf{\%} \\
			& \SetCell[c=3,r=1]{l} \textbf{Example} & & \\
			\midrule[\heavyrulewidth]
			1 & Delete a slash (/) once. & 49,169 & 10.56 \\
			& \SetCell[c=3,r=1]{l} \red{/}s12445-012-0033-7 $\longrightarrow$  s12445-012-0033-7  & & \\
			\midrule
			2 & Add a hyphen (-) four times. & 21,918 & 4.71 \\
			& \SetCell[c=3,r=1]{l} 9781591401087.ch001 $\longrightarrow$ 978\blue{-}1\blue{-}59140\blue{-}108\blue{-}7.ch001 & &  \\
			\midrule
			3 & Delete ``2'' twice, add ``5'' once, add ``7'' once, & 18,607 & 3.99 \\
			& replace ``8'' with ``9'' once, and replace ``6'' with ``3'' once. &  &  \\
			& \SetCell[c=3,r=1]{l} \red{22}14-\purple{8}\purple{6}47\_dnp\_e1000010 $\longrightarrow$ 1\blue{57}4-\purple{9}\purple{3}47\_dnp\_e1000010 & & \\
			\bottomrule
		\end{booktabs}
	\end{center}
\end{table}

\cref{tbl:diff_list_only_suffix_changes} presents frequent patterns for ``only the suffix changed.'' The most frequent pattern was the correction of ``//'' to ``/'' as a separator between the prefix and the suffix (10.56\%). The second most frequent pattern was the correction of the ISBN in the suffix by adding hyphens as separators (4.71\%). The third most frequent pattern was the replacement of the ISSN in the suffix with a new one (3.99\%). According to these results, we must be careful not to set double slashes between the prefix and the suffix. When we apply other identifiers such as the ISBN and ISSN to suffixes, we may need to format or update them owing to changes in the ISBN or ISSN.

\section{Conclusion}

We identified and analyzed the deleted DOIs. As a result, we identified 708,282 deleted DOIs that existed in March 2017 and did not exist in January 2021 using the proposed method. The majority of these DOIs were individual scholarly articles such as journal articles, proceedings articles, and book chapters. The cases where many DOIs were assigned to the same content were retracted papers in a specific volume of a journal or abstracts of international conference proceedings. We revealed the factors that caused a large number of deleted DOIs. The findings of this study are useful for both considering the problems caused by deleted DOIs in citation analysis and altmetrics and assigning DOIs in a better way to avoid deleted DOIs.

The limitations of this study are outlined below, and directions for future work are suggested. First, we were unable to identify deleted DOIs that existed before March 2017, i.e., DOIs that had been deleted as of March 2017. We were also unable to capture the deleted DOIs after March 2021. 
Other DOIs whose contents were not reachable were beyond the scope of this study. The deleted DOIs in this study were defined only as deletions of identifiers and metadata. Therefore, we will expand the scope of the deleted DOIs and develop a methodology for identifying them beyond this definition. Second, we were unable and classify all the factors that caused deleted DOIs and classify them, because we only focused on the most frequent cases. Moreover, it is unclear whether the observed factors represented exceptional cases. Third, a quantitative analysis of the effects of deleted DOIs on citation analysis and altmetrics is important, e.g., examining whether Alias/Primary DOIs are included in the citation indexes and altmetrics. Recently, DOI assignments to the contents on preprint servers such as arXiv have been actively conducted~\cite{arxiv_blog}, and there is a concern that the numbers of duplicated DOIs and deleted DOIs will increase. Analyzing these DOIs is beyond the scope of this study. However, it is important to understand the stability and persistence of the DOI system in the future. The dataset used in this study is available at Zenodo~\cite{Kikkawa_dataset_tpdl2022}.

\vspace{-0.4cm}
\subsubsection*{Acknowledgments.}
This work was partially supported by JSPS KAKENHI Grant Numbers JP21K21303, JP22K18147, JP20K12543, and JP21K12592. We would like to thank Editage (\url{https://www.editage.com/}) for the English language editing.

\newpage{}
\bibliographystyle{splncs04}
\bibliography{reference}

\end{document}